\documentclass[aps,prx,twocolumn,superscriptaddress]{revtex4-2}
\usepackage{orcidlink}
\usepackage{graphics, graphicx}
\usepackage{siunitx}
\usepackage{xcolor}

\newcommand{\LENS}{European Laboratory for Nonlinear Spectroscopy (LENS), University of Florence, 50019 Sesto Fiorentino, Italy}
\newcommand{\CNRINOSesto}{Istituto Nazionale di Ottica del Consiglio Nazionale delle Ricerche (CNR-INO) c/o LENS, 50019 Sesto Fiorentino, Italy}
\newcommand{\DepPhysSesto}{Department of Physics, University of Florence, 50019 Sesto Fiorentino, Italy}
\newcommand{\INRIM}{National Institute for Metrological Research (INRiM), Sesto Fiorentino Office - c/o LENS, Via Nello Carrara, 1 - 50019 Sesto Fiorentino, Italy}

\begin{document}
\title{Parallel analog quantum simulation in homogeneous quantum gases}

\author{D.~Hern\'andez-Rajkov~\orcidlink{0009-0002-1908-4227}}
\email[e-mail: ]{rajkov@lens.unifi.it}
\affiliation{\CNRINOSesto}
\affiliation{\LENS}

\author{A.~Terenzi~\orcidlink{0009-0000-9463-3463}}
\affiliation{\LENS}
\affiliation{\CNRINOSesto}

\author{M.~Fr\'ometa Fern\'andez~\orcidlink{0000-0003-4937-8306}}
\affiliation{\CNRINOSesto}
\affiliation{\LENS}

\author{N.~Grani~\orcidlink{0000-0001-6107-9726}}
\affiliation{\CNRINOSesto}
\affiliation{\LENS}
\affiliation{\DepPhysSesto}

\author{M.~Inguscio \orcidlink{0000-0001-8152-8103}}
\affiliation{\CNRINOSesto}
\affiliation{\LENS}

\author{G.~Del Pace~\orcidlink{0000-0002-0882-2143}}
\affiliation{\INRIM}
\affiliation{\LENS}

\author{G.~Roati~\orcidlink{0000-0001-8749-5621}}
\affiliation{\CNRINOSesto}
\affiliation{\LENS}

\date{\today}

\begin{abstract}
Analog quantum simulation offers a powerful way to study strongly correlated quantum systems that are beyond the reach of classical computation. In this context, ultracold atomic gases have been demonstrated to be an exceptionally versatile and well-controlled platform for implementing various quantum Hamiltonians. 
In this work, we extend this level of control to a multiplexed configuration in which distinct quantum-simulation units are independently controlled and engineered starting from a single atomic cloud. We demonstrate multiplexed operation in two representative settings. 
First, by shaping box-trap potentials and separately controlling the evaporative cooling trajectories, we prepare subsystems at various temperatures across the superfluid transition of the unitary Fermi gas. Second, we demonstrate parallel quantum simulation of the Josephson Hamiltonian across distinct Josephson-junction quantum simulation units with individually tunable parameters, including local phase control to initialize the dynamics. Our scheme provides a versatile route toward systematic studies of dynamics and transport Hamiltonians in strongly correlated ultracold matter. Moreover, it is readily extendable to a wide range of atomic species, geometries, and dimensionalities.
\end{abstract}
\maketitle

% ============================================================
\section{Introduction}
% ============================================================

Understanding the emergent behavior of interacting quantum many-body systems remains a central challenge in modern physics. Macroscopic phenomena such as quantum phase coherence, superfluidity, and non-equilibrium transport emerge from microscopic interactions~\cite{FetterWalecka2012}. Predicting their behavior is extremely challenging. The computational resources required to describe these systems grow exponentially with the number of particles, rapidly exceeding the capabilities of classical theoretical and numerical methods. Quantum simulation offers a powerful approach to realize many-body Hamiltonians under highly controllable platforms, enabling direct investigation of their real-time dynamics~\cite{Georgescu}.

Among the available platforms, ultracold atomic gases have emerged as especially versatile due to their exceptional degree of control over their interactions, geometry, and external potentials~\cite{doyon_generalized_2025, brandstetter_emergent_2025, gross_quantum_2017}. Within these systems, homogeneous quantum gases confined in structured optical box potentials provide a natural platform for realizing bulk many-body quantum matter~\cite{navon_quantum_2021}. 
They provide an ideal setting for studies of quantum transport Hamiltonians, critical phenomena, and collective excitations across interaction regimes in both bosonic and fermionic systems. 
They have enabled non-equilibrium studies through spatially resolved access to local observables such as density, temperature, condensate, and superfluid fractions and multi-order correlation functions via single atom detection in continuum~\cite{gaunt_bose-einstein_2013, mukherjee_homogeneous_2017, li_observation_2024,Verstraten2025}.
Atomic quantum simulators typically probe a single point in a many-body parameter space per experimental cycle. As a result, exploring large parameter spaces is susceptible to slow, unavoidable experimental drifts. 
Progress in these systems has focused on reducing experimental cycle times~\cite{hammel_modular_2025, jain_programmable_2025, vendeiro_machine-learning-accelerated_2022, otoishi_rapid_2020, hetzel_all-optical_2025}, including continuous Bose–Einstein condensate production~\cite{chen_continuous_2022} and modular quantum gas machines that reach sub-second repetition rates~\cite{hammel_modular_2025}. However, these approaches remain fundamentally limited by the intrinsic timescales required for preparation sequences such as trap loading and evaporative cooling. 

Parallel operation offers a qualitatively different strategy: instead of accelerating single experimental sequences, it replaces repetition with the simultaneous realization of multiple independent bulk systems within the same experimental run~\cite{huang_two_dimensional_2024}. 
Thus, increasing the information extracted per cycle without modifying the underlying preparation protocol by enabling several measurements per cycle, analogous to Monte Carlo sampling strategies in which many configurations are generated in parallel to improve efficiency. Although parallelization has already been naturally incorporated into atom tweezers and Rydberg-based atom platforms for quantum computation and simulation~\cite{bluvstein_fault-tolerant_2026, bluvstein_logical_2024, evered_high-fidelity_2023, schine_long-lived_2022, young_half-minute-scale_2020, eckner_realizing_2023}, similar strategies have not yet been realized for bulk quantum gases.  Implementing such a multiplexed framework in homogeneous quantum gases, while preserving homogeneity, coherence, and independence between subsystems still remains an open challenge.

\begin{figure}[h!]
    \centering
    \includegraphics[width=\linewidth]{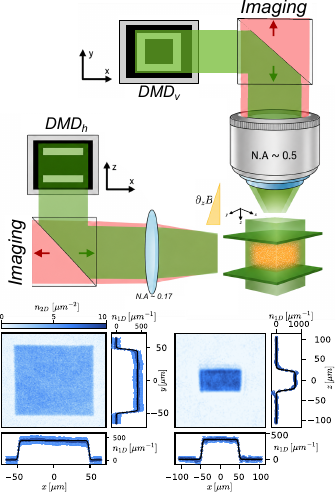}
    \caption{
    \textbf{Experimental setup for production of homogeneous Fermi gases. }
    The atomic gas is confined by optically repulsive potentials generated with two DMDs, projected along the vertical (DMD$_v$) and horizontal (DMD$_h$) directions, which also define our imaging axes. A magnetic-field gradient levitates the gas, compensating gravity for both hyperfine states. 
    Bottom panels: representative column-density images of a homogeneous gas at unitarity, acquired along both imaging axes, for a system of dimensions $100_{(x)}\times100_{(y)}\times50_{(z)}\,\mu\mathrm{m}^3$. Side panels show the integrated 1D profiles along each axis, black curve correspond to the ideal homogeneous profile.
    }
    \label{fig0}
\end{figure}

In this work, we demonstrate a scalable approach to parallel analog quantum simulation using three-dimensional homogeneous ultracold Fermi gases. By dynamically reshaping configurable box-like optical potentials, we realize multiple independent, and homogeneous, mesoscopic quantum-simulation units (QSUs) within a single experimental cycle, each with locally tunable density, temperature, and potential landscape. Synchronous readout across all QSUs enables ensemble-style measurements and direct comparisons between replicas, increasing the experimental throughput with the number of parallel units without requiring additional hardware modifications. We establish the capabilities of the platform in two paradigmatic settings. First, we implement local temperature control and perform simultaneous thermometric characterization of QSUs spanning the normal-to-superfluid transition. Second, we realize a nine-site array of independent QSUs realizing atomic Josephson junctions with tunable tunneling, probing simultaneously the weak-link to tunneling crossover. The dynamics are triggered by locally imposing a prescribed relative phase on each reservoir pair, demonstrating independent phase control across QSUs. Our results establish an experimental framework for multiplexed analog quantum simulation in homogeneous gases, providing a direct realization of parallel operation and opening a route toward efficient exploration of many-body parameter spaces.

% ------------------------------------------------------------
\section{Multiplexing continuum systems}
% ------------------------------------------------------------

\begin{figure*}[ht!]
    \centering
    \includegraphics[width=\linewidth]{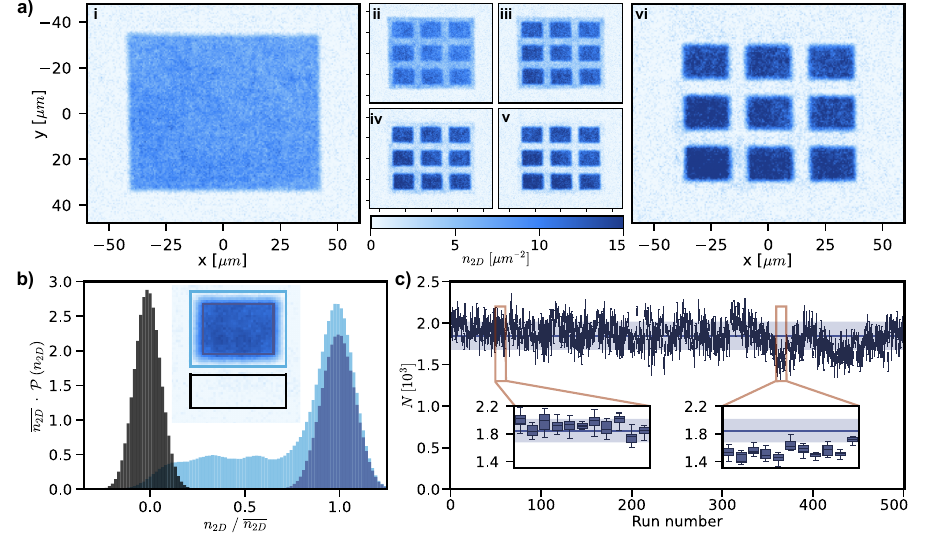}
        \caption{\textbf{Similarity analysis of QSU arrays for unitary Fermi gases.}
        \textbf{a)} Procedure for multiplexing continuum systems, starting from an initial system of size $70_{(x)}\times85_{(y)}\times20_{(z)}\,\mu\mathrm{m}^3$. Representative in situ images are shown at $t=0$ (i), 12 (ii), 24 (iii), 36 (iv), 48 (v), and 100 (vi) ms. The initial system is partitioned into smaller rectangular QSUs following the procedure described in the main text. Images are averaged over 5 experimental repetitions.
        \textbf{b)} Measured probability density $\mathcal{P}(n_{2D})$ of the column-integrated density $n_{2D}$, where $\overline{n_{2D}}$ denotes the average bulk density. Histogram colors indicate the regions from which $n_{2D}$ is sampled.
        \textbf{c)} Box plot showing the minimum, quartiles, and maximum atom number in the corner subsystems as a function of experimental repetition. The mean atom number of the full distribution is shown as a horizontal line, with the corresponding standard deviation indicated by the shaded region.
        }
    \label{fig1}
\end{figure*}

We produce balanced Fermi superfluids using the lowest and third-lowest hyperfine states of $^{6}$Li in the vicinity of their broad Feshbach resonance at $690\,\mathrm{G}$. By tuning the interatomic s-wave scattering length $a$, we access interaction regimes spanning the molecular Bose-Einstein condensate (BEC) to Bardeen-Cooper-Schrieffer crossover~\cite{Hernandez2024, Fernandez2025}. The atomic gas is initially cooled in a crossed optical dipole trap by forced evaporative cooling near resonance. The gas is then adiabatically transferred into a dark optical box potential formed by the superposition of two independently engineered repulsive optical potentials generated by a pair of Digital Micromirror Devices (DMDs). One, DMD$_h$, provides vertical confinement, while the other, DMD$_v$, defines the in-plane geometry, see Fig.~\ref{fig0} and further description of Appendix A. The dual-DMD configuration provides substantial experimental flexibility, enabling real-time control of both the confinement geometry and the potential landscape with refresh rates of up to 17 kHz and support for dynamic sequences comprising up to $3 \times 10^4$ frames per DMD. From the optical dipole trap, we transfer up to 80\% of the atoms into the box potential, the efficiency of this step is primarily limited by the finite spatial extent of the DMDs, and by the temperature of the cloud in the optical dipole trap. The resulting homogeneous sample generally contains $N\approx 5\times10^4$ atoms per state, and remain confined to a single \textit{large} box potential, see Fig.~\ref{fig0} and Fig.~\ref{fig1}a-i. 
For typical atomic densities of $n_{3D} \sim 0.4\,\mu\mathrm{m}^{-3}$ per state, we obtain $k_F = (6\pi^2 n_{3D})^{1/3} \sim 2.8\,\mu\mathrm{m}^{-1}$, and Fermi energy $E_F = \hbar^2 k_F^2/2m \approx 7\,\mathrm{kHz}\times h$. After initial preparation, the gas temperature is typically $T < 0.2\,T_F$, where the Fermi temperature is $k_B T_F = E_F$.

We realize arrays of QSUs by dynamically reshaping the optical potential over a span of hundred's of milliseconds, transforming the initial box into an array of smaller box potentials. 
In Fig.~\ref{fig1}a, we show an example of such transformation, where each QSU has a horizontal extent $20_{(x)} \times 16_{(y)}\,\mu\mathrm{m}^2$ and is separated from its neighbors by $8\,\mu\mathrm{m}\approx 22k_F^{-1}$, ensuring full independence throughout QSUs. The separation procedure consists of gradually introducing repulsive optical barriers between neighboring regions using DMD$_v$, as illustrated in Fig.~\ref{fig1}a ii-iv. As the barriers height increases, atoms in the separating regions are expelled and redistributed into the adjacent dark regions, increasing the atom density of each QSU. To ensure uniformity of atom number across QSUs, the barrier geometry is carefully designed. We construct the boundaries around each QSU using a generalized Voronoi approach: rather than computing cells around point seeds, we compute interfaces around each QSU boundaries, adjusting these to ensure that the marginal area between QSUs is equalized. Under the assumption of an initially uniform density distribution, this geometric design ensures that each barrier region expels approximately equal numbers of atoms to each QSU as the potentials are raised, with residual corrections arising from density gradients in the initial cloud. Conversely, QSUs with distinct densities can be readily realized by tuning the separation process, enabling systematic comparative experiments as a function of density.

Because the ramp duration greatly exceeds the characteristic many-body time scale $h/E_F\sim\ 0.15\,\mathrm{ms}$, atom redistribution proceeds quasi-adiabatically, minimizing residual excitations and density waves generated during the splitting. At the end of the process, the atoms are allowed to equilibrate for an additional $100\,\mathrm{ms}$, after which an evaporation stage is applied to achieve the target degeneracy. Simultaneous evaporation is realized reducing the incident power to the DMD$_h$ without fully compensating gravity, thus allowing atoms to escape through the lower vertical boundary.

The density homogeneity within each QSU is verified by measuring the distribution of the column densities $n_{2D}$ extracted from \textit{in situ} images. As expected for a box potential, the probability distribution $P(n_{2D})$ is narrow and sharply peaks around the mean $\overline{n_{2D}}$, see Fig.~\ref{fig1}b, confirming that each subsystem experiences a nearly homogeneous local environment. To demonstrate similarities among QSUs, Fig.~\ref{fig1}c shows a box plot summarizing the minimum, quartiles, and maximum atom numbers across all replicas over multiple experimental runs. 
Following this multiplexing protocol, atom-number fluctuations between replicas are suppressed below the shot-to-shot fluctuation level, reproducing a small canonical ensemble, even in the presence of large drifts in the overall atom number, see inset near the run number 350. Synchronous data acquisition across all QSUs thus naturally mitigates slow experimental drifts, removing a systematic limitation that typically constrains precision in ultracold-atom experiments. 

\begin{figure*}[ht!]
    \centering
    \includegraphics[width=\linewidth]{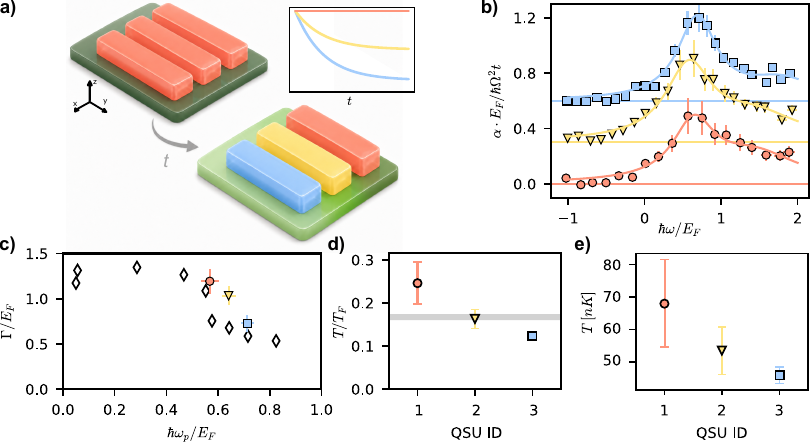}
        \caption{\textbf{Control of local temperature and degeneracy of Fermi gases at unitarity.}
        \textbf{a)} Local evaporative cooling implemented by dynamically engineering the DMD$_h$ pattern intensity over time.
        \textbf{b)} Measured RF response for each QSU, obtained using a $t=4\,\mathrm{ms}$ pulse. Here, $\omega=0$ corresponds to the bare atomic transition. The continuous lines are a guide to the eye.
        \textbf{c)} Full width at half maximum $\Gamma$ as a function of the peak position $\hbar \omega_p$ of the spectral response. Calibration data from Ref.~\cite{mukherjee_spectral_2019} are shown as open symbols.
        \textbf{d)} Extracted $T/T_F$ obtained following the RF thermometry protocols of Refs.~\cite{mukherjee_spectral_2019, yan_thermography_2024, ji_observation_2024}. The gray band indicates the expected critical temperature for superfluidity, $T_c=0.16(1)\,T_F$.
        \textbf{e)} Absolute temperature $T$ of the QSUs.
        }
    \label{fig2}
\end{figure*}

% ------------------------------------------------------------
\section{Local temperature control}
% ------------------------------------------------------------

As a first demonstration of the multiplexed capabilities of our platform, we show that the temperature and quantum degeneracy of each individual QSU can be controlled independently. Since $T/T_F \sim T\,N^{-2/3}$, both quantities can be tuned by controlling the atom number and temperature. We first prepare a spin-balanced unitary Fermi gas in a single large box at $T/T_F \sim 0.25$, above the critical temperature for superfluidity, and subsequently partition it into three QSUs as described in the previous section. Leveraging the dynamic control provided by DMD$_h$, we perform independent evaporative cooling ramps on each QSU, locally lowering each trap depth over $500\,\mathrm{ms}$, see Fig.~\ref{fig2}a. Evaporation proceeds predominantly in the vertical direction, resulting in spatially selective atom removal with negligible inter-subsystem crosstalk. This procedure enables independent control of both the evaporation rate and the final temperature in each subsystem, see Fig.~\ref{fig2}a.

Partitioning the atomic cloud limits the access to global observables, including the momentum distribution, which is typically measured after time-of-flight expansion and used to extract temperatures from the thermal tails~\cite{li_second_2022, gaunt_bose-einstein_2013, ji_observation_2024}. We therefore implement local \textit{in situ} thermometry based on radio-frequency (RF) spectroscopy following the approach of Refs.~\cite{mukherjee_spectral_2019, yan_thermography_2024}. In a unitary Fermi gas, the RF response exhibits a quantitative temperature dependence: both the spectral peak position and the full width at half maximum provide sensitive thermometric observables. Hence, extraction the local reduced temperature $T/T_F$ even in the absence of global thermal equilibrium is possible~\cite{yan_thermography_2024}. Crucially, global RF excitation combined with spatially resolved detection allows simultaneous acquisition of RF spectra across all units.

\begin{figure*}[ht!]
    \centering
    \includegraphics[width=\linewidth]{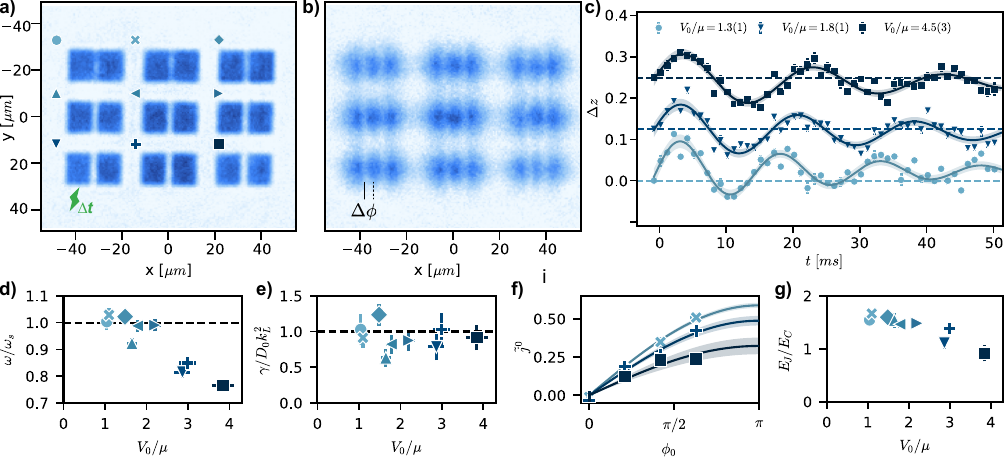}
        \caption{\textbf{Parallel probing of the weak-link to tunneling crossover.}
        Experimental realization of nine independent QSUs implementing atomic Josephson junctions in the molecular BEC regime. \textbf{a)} In situ and \textbf{b)} time-of-flight images after 2 ms of expansion. Both panels show averages over 5 experimental repetitions. Each QSU implements a different tunneling barrier, with height $V_0/\mu$ ranging from 1 to 4 specified by the symbols.
        \textbf{c)} Time evolution of the imbalance $\Delta z$ after imprinting an initial phase difference $\phi_0=0.4\pi$, shown for selected values of $V_0/\mu$. Measurements are fitted with the damped-oscillation function $Ae^{-\gamma t/2}\sin\left(\sqrt{\omega^2-(\gamma/2)^2}\,t+\varphi\right)$, solid lines show the best fits, and shaded regions indicate the corresponding one-standard-deviation fit uncertainty.
        \textbf{d)} Oscillation frequency $\omega$, normalized by the speed-of-sound frequency $\omega_s$.
        \textbf{e)} Damping rate $\gamma$, normalized by the diffusive damping scale $D_0k_L^2$.
        \textbf{f)} Initial Josephson current as a function of the imprinted phase $\phi_0$. Lines show sinusoidal fits to the current--phase relation, $\tilde{j}^0=\tilde{j}_c\sin\phi_0$.
        \textbf{g)} Ratio $E_J/E_C$ as a function of $V_0/\mu$.
        }
        
    \label{fig3}
\end{figure*}

To probe the RF response, we employ a compact shape-optimized RF coil designed to maximize coupling to the atomic magnetic dipole~\cite{Scazza2025}. The RF field drives the transitions between the hyperfine states $\vert 1\rangle = \vert ^2\mathrm{S}_{1/2}\,, m_J=-1/2\,, m_I=1\rangle$ and $\vert 2\rangle = \vert ^2\mathrm{S}_{1/2}\,, m_J=-1/2\,, m_I=0\rangle$, the latter being initially unoccupied. The spectral response is characterized by the transferred fraction $\alpha(\omega)=N_{\vert 2\rangle}/N_{\vert 1\rangle}^{0}$ as a function of the RF detuning $\omega=\omega_{\mathrm{RF}}-\omega_{\vert 1\rangle\rightarrow\vert 2\rangle}$, where $N_{\vert 2\rangle}$ denotes the number of atoms transferred to state $\vert 2\rangle$ and $N_{\vert 1\rangle}^{0}$ the initial population in state $\vert 1\rangle$. In the Fermi's golden rule regime~\cite{chen_emergence_2025}, the corresponding normalized spectral function is given by $I(\omega)=\alpha(\omega)E_F/(\hbar\Omega^2 t)$ where $\Omega$ is the Rabi frequency and $t$ the duration of the RF pulse. The Rabi frequency is calibrated via Rabi oscillations in a fully spin-polarized gas prepared in state $\vert 1\rangle$, which yields $\Omega\simeq2\pi\times504\,\mathrm{Hz}$. We employ RF pulses of duration $t=4\,\mathrm{ms}$, ensuring operation in the linear-response regime~\cite{chen_emergence_2025}.
Figure~\ref{fig2}b shows the measured spectral response as a function of the RF detuning for all QSUs.

Across QSUs, the spectral profiles exhibit systematic broadening and a shift of the peak position, indicative of the distinct thermal states. To convert these spectral observables into reduced temperatures $T/T_F$, we compare the measured spectra to the calibrated reference spectra for homogeneous unitary Fermi gas~\cite{mukherjee_spectral_2019} that provide a standard calibration dataset over a wide range of degeneracy. The agreement between the measured values and the calibration curve shown in Fig.~\ref{fig2}c corroborates the thermometric method and supports the assumption that each unit is locally in thermal equilibrium. The observed uncertainties primarily reflect systematic effects linked to the low Fermi energy of the system, which with increased atom number would offer an improved quantitative precision.
Nonetheless, as shown in Fig.~\ref{fig2}d, we can realize three physically distinct regimes: above, near, and below the superfluid critical temperature $T_c \sim 0.16\,T_F$. 
In addition, we demonstrate the independent tuning of the absolute temperature $T$ across subsystems. The Fermi temperature $T_F$ is extracted directly from the measured density of each uniform subsystem, providing a determination of $T$, see Fig.~\ref{fig2}e.
Our scheme allows the simultaneous interrogation of different many-body phases within a single experimental run, providing the possibility of direct comparisons between normal and superfluid regimes. 

% ------------------------------------------------------------
\section{Parallel data acquisition of atomic Josephson Junctions}
% ------------------------------------------------------------

We demonstrate analog parallel quantum simulation within the QSU architecture by focusing on the implementation of the Josephson Hamiltonian. To this end, we prepare nine Josephson junctions (JJs) on the BEC side of the Feshbach resonance at $1/k_Fa\sim2$.
The atomic Josephson junctions are realized by projecting repulsive optical barriers using the DMD$_v$~\cite{kwon_strongly_2020, del_pace_tunneling_2021}, see Fig.~\ref{fig3}a. All optical barriers have a Gaussian $1/e^2$ beam waist $w=0.41(2)\,\mu\mathrm{m}$, comparable to the healing length of the condensate $\xi=1/\sqrt{8\pi n_{3D}a_M}\sim0.9(1)\,\mu\mathrm{m}$, which yields $w/\xi\sim0.5(1)$, where $a_M=0.6\,a$ is the molecular scattering length. Each junction is confined in box traps of volume $30_{(x)}\times14_{(y)}\times14_{(z)}\,\mu\mathrm{m}^3$, and kept at a temperature $T<0.4\,T_c$, with a chemical potential $\mu/h \sim 480\,\mathrm{Hz}$. Residual potential imperfections remain below $8\,\mathrm{Hz}$, much smaller than $\mu$, corresponding to density inhomogeneities below $2\%$, see Appendix A.

We independently tune the tunneling energy per particle $E_J=\hbar I_c$ by varying the dimensionless barrier height $V_0/\mu\in[1,4]$, and where $I_c$ is the critical super-current. In general, the relevant control parameter in JJs is the ratio $E_J/E_C$, where $E_C=2\left(\partial\mu/\partial N\right)_V$ is the charging energy. In the BEC regime, where $\mu\approx gn$, one obtains that $E_C\approx4g/V$ is independent of the number of atoms and instead is solely determined by the interaction strength $g=4\pi\hbar^2a_M/M$, with $M=2m_{^6\mathrm{Li}}$ and the volume of the system~\cite{zaccanti_critical_2019}. Initially, each QSU junction is prepared with zero population imbalance, $\Delta z = (N_L-N_R)/(N_L+N_R)\sim 0$, where $N_L$ and $N_R$ denote the atom numbers in the left and right reservoirs respectively. 

The dynamics are initiated by imprinting a relative phase $\Delta\phi=\phi_0$ onto one of the reservoirs of each QSU~\cite{luick_ideal_2020}. This is achieved by applying, with DMD$_v$, a spatially uniform optical potential matching the shape of the left reservoir for a short duration of $\Delta t=100\,\mu\mathrm{s}\approx0.05\,h/\mu$, sketched in Fig.~\ref{fig3}a. On this timescale, the potential imprints a relative phase $\phi_0\propto\Delta t$ while weakly perturbing the density. The imprinted phase is proportional to the intensity of the applied potential. By independently adjusting the DMD$_v$ pattern for each QSU, the initial phase $\phi_0$ can be tailored individually. In the present work, however, all QSUs are prepared with the same initial phase $\phi_0$, ensuring that differences in their dynamics arise solely from the different barrier tunneling energies.

We monitor the dynamics by measuring each QSU population imbalance $\Delta z$ from \textit{in situ} images, and the relative phase $\Delta\phi$ from the phase of the matter wave interference pattern emerging after the short time-of-flight, see Fig.~\ref{fig3}b. The limited space between neighboring QSUs constrains the time-of-flight up to $2\,\mathrm{ms}$, which is nevertheless sufficient to resolve the matter wave interference pattern between reservoirs of each QSU, while avoiding overlap between adjacent units. Figure~\ref{fig3}c display the behavior of $\Delta z$ as a function of time for selected tunneling junctions, from these, we extract the oscillation frequency and damping rate for each QSU.

In the weak-link regime, the oscillation frequency is set by the sound mode $\omega_s=\pi c_s/L_x$~\cite{luick_ideal_2020,singh_weak-link_2025}, while in the tunneling regime it is governed by the Josephson plasma frequency, $\hbar\omega_J=\sqrt{E_JE_C}$, with $\omega_J<\omega_s$. As shown in Fig.~\ref{fig3}d, the oscillation frequency decreases monotonically with increasing $V_0/\mu$, consistent with the crossover between the two regimes. In particular, coherent oscillations remain clearly visible even for $V_0/\mu\sim4$, reflecting the strong spatial overlap of the two condensates enabled by the condition $w\sim\xi$. In contrast, in this case, the damping rate is nearly independent of the barrier height in the explored range (Fig.~\ref{fig3}e) and is consistent with a hydrodynamic diffusion rate $D_0k_L^2$, where $k_L=\pi/L_x$ and $D_0=\hbar/M$ the quantum diffusivity~\cite{patel_universal_2020,hilker_first_2022, Fernandez2025, NoteDelpace2026}.

To compare junctions with different parameters, we define the dimensionless critical current density $\tilde{j}_c$ via $\tilde{j}_c = MI_c A_\perp / \hbar$, with $A_\perp = L_yL_z$ the JJ's cross-section area. The critical current density $\tilde{j}_c$ for each QSU is extracted from the measured initial current density $\tilde{j}_c^0$ using the current-phase relation expected at $t=0$, $\tilde{j}_c^0 = \tilde{j}_c \sin\phi_0$, as a function of the imprinted phase $\phi_0$, see Fig.~\ref{fig3}f. The maximum value of the current-phase relation, $\tilde{j}_c$, decreases monotonically with increasing $V_0/\mu$, reflecting the progressive suppression of quantum tunneling. The ratio $E_J/E_C$ correspondingly decreases by approximately a factor of two over the explored parameter regime reaching unity for the largest $V_0/\mu$, see Fig.~\ref{fig3}g. Even lower values of $E_J/E_C$ can be accessed by simultaneously increasing $V_0/\mu$, and by reducing the trap volume to enhance $E_C$. 

Our results thus pave the way toward controlled investigations of the regime $E_J\sim E_C$, where quantum phase fluctuations are expected to become dominant. Here, the parallel operation of many junctions in different regimes would allow for the unambiguous assessment of the emergence of their distinctive quantum behavior. By keeping a portion of QSU junctions operating in the classical regime as a reference, the genuine quantum effects exhibited by the remaining ones could be isolated from general experimental fluctuations commonly affecting all junctions. 

\section{Conclusions}
We have introduced a scalable framework for parallel analog quantum simulation using three-dimensional homogeneous ultracold Fermi gases. Our protocol demonstrates the simultaneous realization of independent mesoscopic subsystems within a single experimental cycle. By dynamically reshaping configurable box-like optical potentials, we show independent control over the density, temperature, quantum phase, and potential landscape of each QSU. 
This approach goes hand in hand with recent advances aimed to reduce the experimental duty cycle through optimized loading and evaporative cooling protocols, and provides a complementary route to increased the experimental throughput~\cite{hammel_modular_2025, jain_programmable_2025, vendeiro_machine-learning-accelerated_2022, otoishi_rapid_2020, hetzel_all-optical_2025}.

Beyond these practical advantages, the multiplexed architecture opens up directions for future research. The ability to prepare QSUs at independently controlled temperatures and degeneracies provides a natural pathway for studying heat and particle transport between reservoirs at different points in the phase diagram~\cite{Washburn1985}, as well as proximity effects between distinct many-body phases, extending the results form the point-contact junctions~\cite{Hausler2021,Huang2025}. 
For example, particle and thermal currents can be driven by temperature or chemical-potential gradients between neighboring subsystems. Such configurations are directly relevant to quantum thermodynamics~\cite{Campbell2025} and to the realization of quantum heat engines with ultracold atomic matter~\cite{Koch2023}. 
More broadly, this platform enables systematic studies of non-equilibrium many-body dynamics across Hamiltonian parameter space and extends naturally to other species, dimensions, and lattice geometries.

% ============================================================================
\section{acknowledgments}
% ============================================================================
We thank the Quantum Gases group at LENS for stimulating discussions, and to T. Braud for participating in the initial stages of this work.
G.R. acknowledge financial support from the PNRR MUR project PE0000023-NQSTI and from the Italian Ministry of University and Research under the PRIN2017 project CEnTraL and project CNR-FOE-LENS-2024. 
The authors acknowledge support from the European Union - NextGenerationEU within the “Integrated Infrastructure Initiative in Photonics and Quantum Sciences" (I-PHOQS). 
This publication has received funding under the Horizon Europe programme HORIZON-CL4-2022-QUANTUM-02-SGA (project PASQuanS2.1, GA no.~101113690) and Horizon 2020 research and innovation programme (GA no.~871124).

% ============================================================================
\section*{Data Availability}
% ============================================================================

The data that support the findings of this study are available from the corresponding author upon reasonable request.

% ============================================================================
\section*{Appendix}
% ============================================================================

\subsection{Box potential}
The optical box is realized by projecting patterned blue-detuned light at $532\,\mathrm{nm}$ from two digital micromirror devices (DMDs) along two orthogonal axes, see Fig~\ref{fig0}. Along the vertical direction, the pattern generated by the DMD$_v$ is imaged onto the atomic cloud through a high-resolution microscope objective~\cite{kwon_strongly_2020}, which is also used for resonant-absorption imaging of the atomic density at $670.9\,\mathrm{nm}$ and for projecting the blue-detuned optical potential~\cite{del_pace_tunneling_2021}. The high numerical aperture of this objective, $\mathrm{NA}\approx0.5$, guarantees sub-micrometrer spatial resolution at both relevant wavelengths~\cite{kwon_strongly_2020} and provides a maximum radial field of view of $200\,\mu\mathrm{m}$. Along the horizontal direction, the pattern from DMD$_h$ is projected onto the atomic cloud using a $4f$ achromatic telescope configuration with numerical aperture $\mathrm{NA}\sim0.17$, corresponding to a diffraction-limited resolution of $\sim2\,\mu\mathrm{m}$; imaging along this axis reuses the same projection telescope. This independent optical control enables precise tuning of both the vertical confinement and the lateral system size, with an in-plane spatial resolution comparable to the superfluid coherence length, which spans $0.5$--$1\,\mu\mathrm{m}$ across a broad range of interaction strengths. 
Gravity is compensated by magnetic levitation, creating an almost homogeneous three-dimensional gas, although a residual in-plane magnetic harmonic potential with trapping frequency $\omega_{xy}\sim2\pi\times7\,\mathrm{Hz}$ introduces a weak overall curvature, while along the $z$ direction an anti-confining magnetic curvature characterized by an imaginary trapping frequency $\omega_z\sim i\,\omega_{xy}$ is also present. Depending on the chemical potential of the interacting gas, these residual curvatures define a finite length scale beyond which density variations become non-negligible, limiting the spatial extent of the homogeneous region. \\
\indent To assess the degree of background potential homogeneity, we consider a system with spatial extension $(d_x,d_y)$ centered at position $(x_0,y_0)$ in the $x$-$y$ plane. We focus on the transverse plane because, in most experiments, the system extends much farther in the $x$ and $y$ directions than along the $z$ axis. The trapping potential at the center of the system is given by $V_0=\frac{1}{2}m\omega_r^2 r_0^2$, while at the system boundary it is $V_{\partial\Omega}=\frac{1}{2}m\omega_r^2 r_{\partial\Omega}^2$, where $\partial\Omega$ denotes the boundary of the system. The resulting potential variation across the system is therefore
$\Delta V = \max\left(V_{\partial\Omega},V_0\right)-\min\left(V_{\partial\Omega},V_0\right)
=\frac{1}{2}m\omega_r^2\left[\max\left(r_{\partial\Omega},r_0\right)^2-\min\left(r_{\partial\Omega},r_0\right)^2\right].$
For the specific parameters considered in Sec.~IV, with the system centered at the trap minimum ($r_0=0$), the potential variation reduces to
$\Delta V=\frac{1}{2}m\omega_r^2\left[\left(\frac{30\,\mu\mathrm{m}}{2}\right)^2+\left(\frac{14\,\mu\mathrm{m}}{2}\right)^2\right]\approx h \times 7.9\,\mathrm{Hz}$.


\begin{thebibliography}{45}
\bibitem [{Not()}]{NoteDelpace2026}%
  \BibitemOpen
  \href@noop {} {}\bibinfo {note} {A detailed investigation of the underlying dissipation mechanisms in the Josephson dynamics is currently being addressed.}\BibitemShut {Stop}%
\end{thebibliography}
\end{document}